\documentclass{pasj00}

\begin{document}
\SetRunningHead{Author(s) in page-head}{Running Head}
\Received{2000/12/31}
\Accepted{2001/01/01}

\title{Suppresion of Self-Phase Modulation in a Laser Transfer System using Optical Fiber on the Subaru Telescope}

\author{Meguru \textsc{Ito},\altaffilmark{1,2}
            Yutaka \textsc{Hayano},\altaffilmark{2}
            Yoshihiko \textsc{Saito},\altaffilmark{3}
            Hideki \textsc{Takami},\altaffilmark{2}
            Norihito \textsc{Saito},\altaffilmark{4}
            Kazuyuki \textsc{Akagawa},\altaffilmark{5}
	   and
	   Masanori \textsc{Iye},\altaffilmark{6}
            }
\altaffiltext{1}{Adaptive Optics Lab, University of Victoria, 3800 Finnerty Road, Victoria, B.C., V8P 5C2, Canada}
\altaffiltext{2}{Subaru Telescope, National Astronomical Observatory of Japan, National Institutes of Natural Sciences, 650 North A'ohoku Place, Hilo, HI 96720-2700, USA}
\email{itomg@uvic.ca}
\altaffiltext{3}{Department of Physics, Graduate School of Science and Engineering, Tokyo Institute of Technology, 2-12-1 H-29, Ookayama, Meguro-ku, Tokyo 152-8550, Japan}
\altaffiltext{4}{Optical Green Technology Research Unit, Advanced Science Institute, RIKEN, 2-1 Hirosawa, Wako, Saitama 351-0198, Japan}
\altaffiltext{5}{Singlemode Co., Ltd., 12-12 Minami-cho, Sakado-shi, Saitama 350-0233, Japan}
\altaffiltext{6}{Division of Optical and Infrared Astronomy, National Astronomical Observatory of Japan, 2-21-1 Oosawa, Mitaka, Tokyo 181-8588, Japan}


\maketitle
\begin{abstract}
We are developing the Laser Guide Star Adaptive Optics (LGS/AO188) system for the Subaru Telescope at Mauna Kea, Hawaii. This system utilizes a combination of an all-solid-state mode-locked sum-frequency generation (SFG) laser (1.7-GHz bandwidth, 0.7-ns pulse width) as a light source and a single-mode optical fiber for beam transference. However, optical fibers induce nonlinear effects, especially self-phase modulation (SPM).  
We studied SPM in our photonic crystal fiber (PCF). SPM broadens the spectrum of a laser beam and decrease the efficiency of bright laser guide star generation. We measured the spectrum width using a spectrum analyzer. We found a spectrum width of 8.4 GHz at full width at half maximum (FWHM). The original FWHM of our laser spectrum was 1.4 GHz. This was equivalent to a 70 \% loss in laser energy.
We also measured the brightness of the sodium cell and evaluated its performance as a function of laser wavelength. The cellfs brightness showed a peculiar tendency; specifically, it did not extinguish even though the wavelength varied by more than 5 pm. 
To reduce the impact of SPM, we developed an optical system that divides one laser pulse into four lower-power pulses. The laser peak power after passing through the new optical system was decreased to one-fourth the original, reducing the impact of SPM on the sodium cell. An actual laser guide star created with the new system was 0.41 mag brighter than the laser guide star created with the original system. 
We achieved brighter laser guide star generation by dividing a laser pulse to reduce its peak intensity. This is an effective method for laser relay using optical fiber.
\end{abstract}

\section{Introduction}\label{sec.intro}

A new laser guide star (LGS) adaptive optics system, called AO188/LGS
, was commissioned at the Nasmyth focus of the Subaru Telescope \citep{Iye04, Takami06, Hayano10}. Our AO188/LGS uses a curvature wavefront sensor  that divides the pupil by a microlens array and feeds the divided beam to 188 fiber-coupled avalanche photo diodes operated in photon counting mode and a bimorph deformable mirror with custom designed 188 driving electrodes. An all-solid-state sum-frequency laser, installed in a temperature controlled laser room on the Nasmyth floor, is used for generating a laser guide star. Optical fiber transmits the laser from the Nasmyth floor to the laser launch telescope (LLT) mounted at the backside of the secondary mirror of the telescope \citep{Hayano06}. The AO188/LGS system of Subaru Telescope is different from AO systems of other 8-10 m class telescopes in the following aspects. The Very Large Telescope (VLT) also uses optical fiber for laser beam transfer but a dye laser is used as the light source\citep{Bonaccini04}. The LGS system on the Keck telescope uses a dye laser and mirrors to relay the laser beam \citep{vanDam06, Wizinowich06}. 

An obvious advantage of using an optical fiber, instead of a mirror train, in relaying the laser beam to the LLT is that it is free from the atmospheric and geometrical disturbances in its optical path. 
On the other hand, high power laser beam transmission through an optical fiber induces nonlinear effects, stimulated Raman scattering (SRS), stimulated Brillouin scattering (SBS), and self-phase modulation (SPM), which can lead to some degradation of the laser quality. SRS and SBS are similar types of scattering leading to a significant reduction of the output power.  These effects are observed when the incident laser power exceeds a given threshold, and limit the amount of power transmitted through the fiber. At power levels beyond this threshold, the output laser power does not increase, even if the incident power is increased. The threshold of these scattering effects depends on the fiber's effective length and is inversely proportional to the square of the mode field diameter (MFD) of the fiber. SBS is not conspicuous for short pulses whose width is less than 1 ns \citep{Agrawal07}. Given that the pulse width of our SFG laser is 0.7 ns, SBS is not a problem. Instead, SRS could be dominant. We chose a photonic crystal fiber (PCF) whose MFD is larger than that for  step index fibers to reduce SRS. We examined both SRS and SBS for our PCF \citep{Ito09}. 

The third nonlinear effect, self-phase modulation (SPM), is a phenomenon that such that an optical fiber induces phase shifts in the pulsed-laser beam. This is caused by the optical Karr effect, observed as a variation in the refractive index of the optical fiber depending on the incident laser pulse power. Due to this effect, the refractive index is slightly higher at the high-intensity pulse center than at low-intensity side lobes of the pulse. This difference in refractive index produces a phase shift in the pulse, leading to a change in the pulse's frequency spectrum. In other words, SPM broadens the laser bandwidth in wavelength. Generally, the broadened pulse width, $\delta\omega_{max}$, can be calculated using the following equation \citep{Agrawal07}:

\begin{equation}
\delta\omega_{max} = 0.86 \gamma L_{eff} P_{0} \Delta \omega \label{eq.spm}
\end{equation}

\noindent
where $L_{eff}$ is the effective length of the fiber, $P_{0}$ and $\Delta\omega$ are the peak power and the pulse width of the incident laser, respectively, and $\gamma$ is a nonlinear coefficient. For our system, $\gamma$ is 3.36 $\times $10$^{-4}$ m$^{-1}$W$^{-1}$, $L_{eff}$ is 33.6 m, $P_{0}$ is 61.9 W, and $\Delta\omega$ is 1.7 GHz. This equation indicates that the broadened pulse width is proportional to the peak power of the laser pulse. If the spectrum of our laser were broadened by SPM, the efficiency of our LGS generation would drop to unacceptably low levels as the number of photons within the wavelength range suitable for exciting sodium atoms is reduced accordingly. In order to avoid this SPM effect, we need to lower the peak pulse power.

Another notable features of SPM is that SPM-induced spectral broadening is accompanied by an oscillatory structure covering the entire frequency range. In general, the spectrum consists of many peaks, ant the outermost peaks are the most intense. The number of peaks depends on $\Delta\omega$ and increases linearly with it \citep{Agrawal07}.

The present paper reports our study  to evaluate the impact of SPM on the brightness of the LGS and to develop a technique for mitigating the SPM effect to achieve high output power for generating a bright LGS.

\section{Experimental Setup}\label{sec.exp}

The specifications of our laser are shown in Table \ref{tbl.laser_spec}. The laser produced $\sim$6.7 W with a bandwidth of about 1.7 GHz tuned to the sodium D$_2$ line at 589.159 nm. Note that the wavelength of our laser is tunable from 589.060 nm to 589.170 nm.

\begin{table}[htbp]
\caption{Specifications of our SFG laser}\label{tbl.laser_spec}
\begin{center}
\begin{tabular}{ll}
\hline \hline
Wavelength & 589.159 nm \\
Output laser power & $\sim$ 6.7 W \\
Oscillation & Active mode-locking \\
\phantom{0} & (Repetition rate: 143 MHz) \\
Bandwidth & 1.7 GHz \\
Pulse width & 0.7 ns \\
Beam quality & TEM$_{00}$, M$^2$ $<$ 1.03 \\
Polarization & linear (horizontal) \\
Power stability & $\pm$ 4 \% (8hrs)\\
\hline 
\end{tabular}
\end{center}
\end{table}

\subsection{One-pulse system}

\begin{figure}[tb]
   \begin{center}
   \begin{tabular}{c}
   \includegraphics[width=8cm]{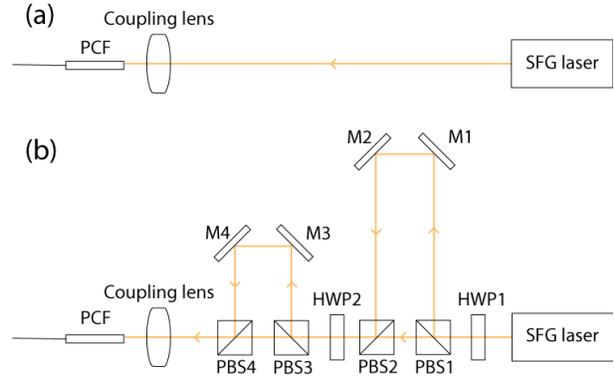} \\
   \end{tabular}
   \end{center}
   \caption{The experimental setup for our SPM measurements. SFG laser: our SFG laser; HWP: half-wave plate; PBS: polarized beam splitter; M: mirror; PCF: photonic crystal fiber. Configuration (a) shows the setup for the one-pulse system, and configuration (b) shows the new optics system, where each pulse is divided into four pulses.}
   \label{fig.exp} 
   \end{figure}

First, we measured the effect of SPM in our original, simple laser coupling optics system as shown in Figure \ref{fig.exp}(a). We input the laser beam to the PCF directly through a coupling lens. We measured the frequency spectrum of the output laser beam using a spectrum analyzer at output powers of 0.02 W, 0.05 W, 0.1 W, 0.51 W, 0.96 W, 1.4 W, 2.0 W, 3.0 W, and 3.7 W. Additionally, we placed a gas cell containing sodium atoms in the path of the output beam and measured output sodium D$_{2}$ emission line intensity profile as a function of wavelength by   modulating the input laser wavelength.

The input mode-locked pulse laser has a 0.7 ns pulse width, a repetition rate of 143 MHz, and a power up to 6.7W.  The laser system has a high  TEM$_{00}$ beam quality, $M^{2} < 1.03$, and stable to $\pm 4\%$ for 8 hrs continuous operation.  Table \ref{tbl.laser_spec} summarizes its characteristics.

Hereafter, we call this original system the "one-pulse system" because the laser pulse was undivided.

\subsection{Four-pulse system}

In order to reduce the peak power to avoid SPM, we installed an optical module that divides the laser pulse into four sub pulses before inputting to the PCF. The new experimental setup is shown in Figure \ref{fig.exp}(b). This system divides the laser beam into four beams by using two modules of a half-wave plate and two polarized beam splitters each. 
Each module diverts half of the beam to an optical delay line and recombined. Therefore a module produces two pulses one of which has a delay corresponding to the optical path length of the delay line.  Since the two delay line modules are designed to give two different pulse lag time, the out beam produces four pulses with 1/4 peak intensity each and three of the resultant pulses have time lags so that the pulse intervals of the four pulses become 1.75 ns, 1/4 of the original pulse interval 7 ns.
The path of Figure \ref{fig.exp}(a) was defined as the basic path. The optical path reflected at PBS1 and recombined at PBS2 was called extension path 1(EP1). The optical path reflected at PBS3 and recombined at PBS4 was termed extension path 2 (EP2). 
At first, the laser pulses were changed their polarization from horizontal to 45$^{\circ}$-linear polarization by the first half-wave plate (HWP1), whose angle was fixed at 22.5$^{\circ}$ from the horizontal polarization. These were divided equally by PBS1. One pulse, divided by its polarization, was propagated along EP1 and was delayed by 3.5 nsec. This corresponded to an extension of 1.05 m. The two recombined pulses were changed their polarization again by HWP2. HWP2 was also fixed at 22.5$^{\circ}$ from the horizontal polarization. PBS3 divided the incident beams equally again. The optical path difference of EP2 was 0.525 m, and it provided 1.75 nsec in separation. This optics system is hereafter called "four-pulse system".

\begin{figure}[tb]
   \begin{center}
   \begin{tabular}{c}
   \includegraphics[width=8cm]{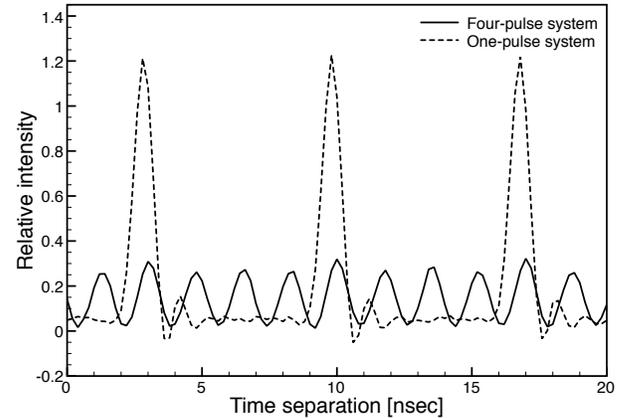} \\
   \end{tabular}
   \end{center}
   \caption{The pulse shape of the laser beam before and after passing through the optics system. The solid line and dashed line represent the divided and ordinal laser pulses, respectively. The peak power of the divided pulse was reduced to one-fourth of the original peak power, as was the separation between pulses.}
   \label{fig.pulse} 
   \end{figure}

Figure \ref{fig.pulse} shows the pulse shape before and after passing through the optics system. The solid line and dashed line represent the divided and original laser pulses. The original laser pulses were separated by 7 nsec, and the separation of the divided pulses was 1.75 nsec. The pulse separation and peak powers were each reduced to about one-fourth of the original values. We measured the variation in the effect of SPM as the laser beam passed through the optics system and the PCF and compared these results with our previous measurements. The spectra were measured at throughput powers of 0.5 W, 1.0 W, 2.1 W, 2.6 W. Additionally, we measured the brightness of a sodium cell in the path of the output laser beam at various wavelengths.

Lastly, we verified the actual difference in the brightness of the LGS produced by the optical configurations with and without the pulse delay modules. We launched the laser into the night sky and measured and compared the brightness of the LGS generated by the one-pulse and four-pulse systems.

\section{Results}\label{sec.res}

\begin{figure}[tb]
   \begin{center}
   \begin{tabular}{c}
   \includegraphics[width=8cm]{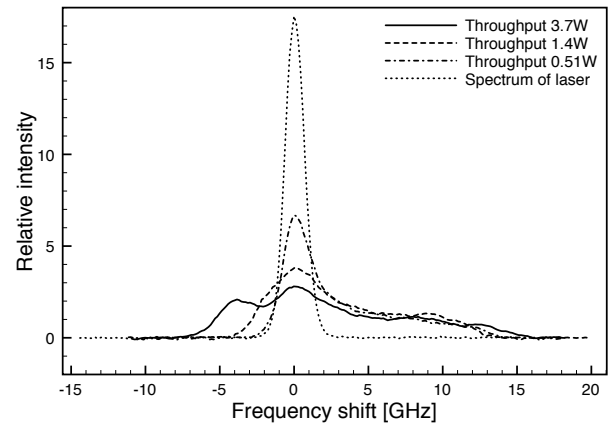} \\
   \end{tabular}
   \end{center}
   \caption{The broadened spectra of the laser beam. The solid line, dashed line, chain line, and dotted line represent the spectrum after passing through the PCF at throughput powers of 3.7 W, 1.4 W, and 0.51 W, and our laser itself, respectively. The areas of each line were normalized to the same value. The measured FWHMs were 1.4 GHz at the laser and 8.4 GHz after passing through the PCF at a throughput power of 3.7 W.}
   \label{fig.spm1} 
   \end{figure}

The spectral profile of the output laser through the PCF is significantly broadened as the input laser power is increased. Figure \ref{fig.spm1} shows the broadened spectrum of the laser beam after passing through the PCF. For comparison, it also shows the laser spectrum (represented by the dotted line) before entering the PCF. The spectra at throughput powers of 3.7 W, 1.4 W, and 0.51 W are represented by the solid line, chain line, and dashed line, respectively. The areas of these spectra were normalized to the same value. The measured full width at half maximum (FWHM) for 0.51 W, 1.4 W, and 3.7 W were 2.2 GHz, 5.1 GHz, and 8.4 GHz, respectively. The FWHM of the spectrum of the laser before passing through the PCF was 1.4 GHz.

These results showed that SPM occurred even though the incident power was very low. Figure \ref{fig.spm1} shows a feature of the measured spectrum that was not symmetric. The SPM broadens the spectrum symmetrically if the incident pulse form is Gaussian. The laser beam spectrum was broadened by the SPM in our system, decreasing the efficiency of creating an LGS at the mesosphere. At 3.7 W, the crossover between the laser spectrum after passing through the PCF and the spectrum before passing through the PCF was only 27 \%. A complete correction for SPM would result in an LGS that is brighter by about 1 magnitude.

\begin{figure}[tb]
   \begin{center}
   \begin{tabular}{c}
   \includegraphics[width=8cm]{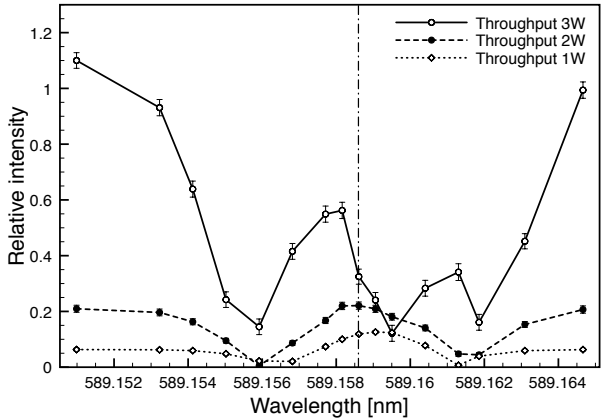} \\
   \end{tabular}
   \end{center}
   \caption{The brightness of the sodium cell, along with the wavelength. The solid line, dashed line, and dotted line represent the brightness at throughput powers of 3 W, 2 W, and 1 W, respectively. The chain line marks 589.1586 nm, the excitation line of the sodium D$_2$ transition. This figure shows an interesting feature of SPM: the number of peaks increased when throughput power increased.}
   \label{fig.cell1} 
   \end{figure}

The brightness of the sodium cell illuminated by the beam after passing through the PCF is shown in Figure \ref{fig.cell1}. The solid line, dashed lines, and dotted line represent throughput powers of 3 W, 2 W, and 1 W, respectively. The chain line marks 589.1586 nm, the excitation line of sodium D$_2$ transition. This result indicates that the influence of SPM was seen even in a sodium cell. The sodium cell was heated to 360 K, which is different from the temperature of sodium layer in the mesosphere (200 K). 
However, the impact of the difference of absorption spectrum of the sodium D$_2$ line between two temperatures would be little because the broadened spectrum of our laser was enough large to cover whole sodium absorption line. Given the difference in temperature, the lack of resemblance between the brightness curves would be consistent with the brightness of the LGS on the sky.
The surprising result in these measurements is that the standard excitation wavelength is not necessarily required to brighten the LGS.

\begin{figure}[tb]
   \begin{center}
   \begin{tabular}{c}
   \includegraphics[width=8cm]{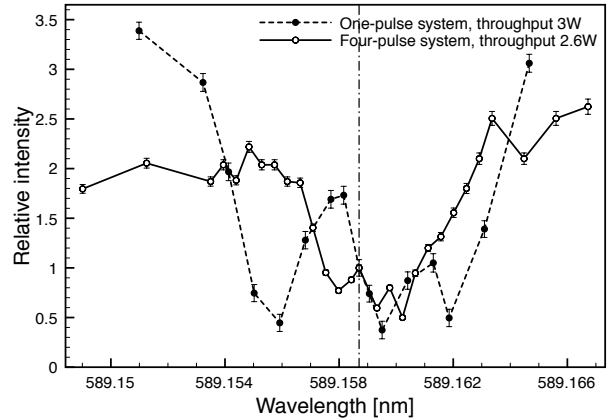} \\
   \end{tabular}
   \end{center}
   \caption{A comparison of the sodium cell's brightness in the one-pulse and four-pulse systems. The solid line and dashed line represented the brightness of the cell for the four-pulse and one-pulse systems, respectively. The chain line marks the sodium atom's excitation wavelength. In the four-pulse system, the throughput power was 2.6 W. In the one-pulse system, the throughput power was 3 W. The line for the one-pulse system was the same as Figure \ref{fig.cell1}. The degree of broadening for four pulses was clearly less than the broadening for one pulse.}
   \label{fig.cell2} 
   \end{figure}

Figure \ref{fig.cell2} provides measured spectra of the one-pulse system with 3 W throughput power and the four-pulse system with 2.6 W. The solid line and dashed line represent the spectra of the four-pulse system and the one-pulse system, respectively. The 589.1586 nm wavelength is indicated by the vertical chain line. The ordinate is normalized by area as described in Figure \ref{fig.cell1}. In the one-pulse system, we note two local maxima, at 589.158 nm and 589.161 nm. 
Theoretically, we expected to observe a peak value higher than these local maxima at the D$_2$ line center. However, we observed a  local maximum slightly offset from the D$_2$ line center and two higher peaks were expected to be at around $\pm$0.01 nm from the wavelength. This is because SPM not only broadens the spectrum but also increases the number of peaks as throughput power increases. Consequently, dividing one pulse into four pulses softened this additional effect of SPM.

\begin{figure}[tb]
   \begin{center}
   \begin{tabular}{c}
   \includegraphics[width=8cm]{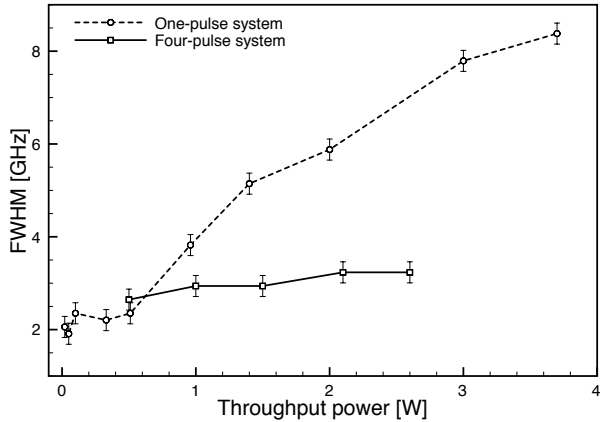} \\
   \end{tabular}
   \end{center}
   \caption{The FWHM of the measured spectra. The solid line and dashed line represent the four-pulse and one-pulse systems, respectively. The FWHM for the four-pulse system grew slowly in comparison with the one-pulse system.}
   \label{fig.fwhm} 
   \end{figure}

The FWHM for each throughput power is shown in Figure \ref{fig.fwhm}. The solid line and dashed line represent the four-pulse and one-pulse systems, respectively. These plots show that the widths of the broadened spectra grew slowly when the pulse was divided into four.

The difference in the brightness of the LGS between the one-pulse and the four-pulse system was 0.41 mag in the R-band. The value was derived to normalize the optical loss and coupling loss. In other words, that shows the difference of the brightness compensated under the condition that launched laser powers were same.

\section{Discussion}\label{sec.dis}

We found that SPM certainly occurred in our fiber relay system, and it probably caused a loss in laser power. Here, we propose a more elaborate numerical evaluation of the spectrum width. The peak power of our laser is 60 W given a throughput power of 3.7 W. The FWHM, in that case, should be 7.8 GHz. Given the asymmetry of the laser spectrum, the FWHM cannot be measured strictly. However the values of the measured FWHM, 8.4 GHz, and the calculated FWHM, 7.8 GHz are close. The limit of the frequency resolution of the spectrum analyzer is 0.15 GHz. If the spectrum was symmetry for the central frequency, that is, the right portion of the spectrum had the same shape as the left portion, the measured FWHM could decrease bringing the value closer to the calculated FWHM. The calculated FWHM was derived with an assumption of symmetric incident pulses.

The asymmetry in the broadened spectrum is caused by the asymmetry in the shape of the laser pulses. The shape of our laser is not considered to be a strict Gaussian. However, given a high peak power and a tight pulse width, the shape of the pulse will approach a Gaussian. As a result, we predict that the broadened spectrum approaches a symmetric shape.

The results of the measurements of the spread in the laser pulse spectrum and the brightness of the sodium cell indicate that dividing the pulse reduced the effect of SPM. However, the addition of optical components increased the loss in laser power in the optical system. The estimated total energy lost in passing through the optical system was 10.3 \% in total. In practice, we measured the loss as 10.8 \%, which is consistent with the calculated value. We compared the energy loss with the increased brightness of the LGS and estimated the efficiency. As described in our results, the brightness of the LGS increased by 0.41 mag under the four-pulse system. That increase was equivalent to a 46 \% increment in the flux of the LGS. 
When the throughput power was 0.96 W for the one-pulse system, an amount of energy absorbed by sodium atoms in the cell was measured as 57.8 \% more than the energy absorbed at a throughput power of 3.7 W. 
When the efficiency of the optics (89.2 \%) and the decrement in the coupling efficiency (about 85 \%), were multiplied by the increment value (57.8 \%), the incremental brightness of the LGS was estimated as 45 \%. 
The coupling efficiency fluctuated slightly day by day; however, the estimated increment in LGS brightness, 45 \%, was consistent with the 46 \% increment in the measured flux. 
Consequently, we concluded that we could generate a brighter LGS by dividing the laser pulse even though the optics system resulted in greater loss in laser power.

Figure \ref{fig.cell2} showed that the excitation efficiency was clearly low at the sodium D$_2$ wavelength under the one-pulse system. However, in the four-pulse system, the peak at this wavelength was slightly lower than the peaks at nearby wavelength. We found that we could generate a brighter LGS when we adjusted the wavelength of the laser by -0.003nm or +0.01nm (relative to the original excitation wavelength) in the four-pulse system. However, the width of the broadened spectrum depends on the coupling efficiency of the PCF. If we can clarify the relationship between the throughput power from the PCF and the width of the laser spectrum, the most appropriate wavelength given the throughput power will be discovered. 
However, we should try to reduce the effect of SPM more effectively to narrow the width of the laser spectrum after the laser beam passes through the PCF.

\section{Summary}\label{sec.sum}

Our laser guide star adaptive optics system uses optical fiber to relay the laser beam. However, some nonlinear effects have to be considered, given the high power transmission through the optical fiber. SRS and SBS, two types of nonlinear scattering, have already been checked in our previous work. These types of scattering do not occur in our system. However, we found that SPM does occur in our PCF. SPM broadened the width of spectrum to 8.4 GHz at maximum laser power. To reduce the impact of SPM, we decreased the peak intensity of the laser pulse. We implemented two optical delay lines modules that divided the laser pulse into four pulses with lower peak intensities. The width of the spectrum after passing through the new optical system and the PCF was 3.2 GHz. We confirmed that SPM had a smaller effect on our new system. Additionally, we found that we could create a LGS that was 46 \% brighter than that created by the original (one-pulse) system. The spectrum of a sodium cell in the four-pulse system also exhibited differences compared with the sodium cell spectrum in the one-pulse system. We found that the brightest LGS is not always generated at the excitation wavelength of the sodium D$_2$ line. In the four-pulse system, the wavelengths that were separated by 0.01 nm from the original excitation wavelength were the most effective. These results will lead to the creation of brighter LGS and to the improved performance of our AO system.

\phantom{00}
We appreciate Dr. Jason Chin and Dr. Kenny Grace of Keck telescope. Their cooperation brings our experiment to a successful conclusion. We thank Mr. Chris Clergeon and Kohei Yamazaki for laser spotters. This work was supported by the Subaru Telescope, which is operated by National Astronomical Observatory of Japan. We are grateful to the staff members of Subaru Telescope for their invaluable assistance in obtaining these data and for their continuous support during IRCS and Subaru AO construction. 


\end{document}